\documentclass[aip, cha,
 amsmath,amssymb,
 reprint,%
]{revtex4-1}

\usepackage{graphicx}% Include figure files
\graphicspath{fig/}
\usepackage{dcolumn}% Align table columns on decimal point
\usepackage{bm}% bold math
\usepackage[utf8]{inputenc}
\usepackage[T1]{fontenc}
\usepackage{mathptmx}
\usepackage{etoolbox}
\usepackage {amsmath}
\usepackage{amssymb}
\usepackage{algorithm}

\usepackage{algorithmic}

\usepackage{newfloat}
\usepackage{listings}
\makeatletter
\def\@email#1#2{%
 \endgroup
 \patchcmd{\titleblock@produce}
  {\frontmatter@RRAPformat}
  {\frontmatter@RRAPformat{\produce@RRAP{*#1\href{mailto:#2}{#2}}}\frontmatter@RRAPformat}
  {}{}
}%
\makeatother
\begin{document}

\preprint{AIP/123-QED}

\title[Cognition Transition]{Transition of AI Models in dependence of noise}
% Force line breaks with \\

\author{Thomas Seidler}
 \altaffiliation[ ]{Physics Department, XYZ University.}%Lines break automatically or can be forced with \\
\author{Markus Abel}%
 \email{markus.abel@ambrosys.de}
\affiliation{ University of Potsdam,
        Institute of Physics and Astronomy,
        Karl-Liebknecht-Str. 24-25,
        14476 Potsdam, Germany.\\
    Ambrosys GmbH, Marlene-Dietrich-Allee 16,
        14482 Potsdam, Germany.
}%

\date{\today}% It is always \today, today,

\begin{abstract}

We investigate the dependence of the score on noise in the data, and on the network size. As a result, we obtain the so-called "cognition transition" from good performance to zero with increasing noise.
    The understanding of this transition is of fundamental scientific and practical interest. We use concepts from statistical mechanics to understand how a changing finite size of models affects the cognition ability under the presence or corrupted data. On one hand, we study if there is a universal aspect in the transition to several models, on the other hand we go into detail how the approach of the cognition transition point can be captured quantitatively. Therefore, we use the so-called scaling approach from statistical mechanics and find a power-law behaviour of the transition width with increasing model size.
    Since our study is aimed at universal aspects we use well-know models and data for image classification. That way we avoid uncertainties in data handling or model setup. The practical implication of our results is a tool to estimate model sizes for a certain "universality class" of models, without the need to investigate large sizes, just by extrapolating the scaling results. In turn, that allows for cost reduction in hyperparameter studies. Here, we present first results on a concrete setup; we think that the understanding the mechanics of large system sizes is of fundamental interest for a further exploration of even larger models.
\end{abstract}

\maketitle

\section*{Lead Paragraph}
\textbf{ We demonstrate empirically that the principles of statistical mechanics hold for deep learning. This is a very deep result, as theoretical physics provides  a well worked-out framework for finite-size effects, scaling theory and universality. This, in turn, has powerful implications for the design of deep networks, and requirements on data quality. In our study, we use well-known data sets and AI models to underline the applicability of our approach. Since it is very important for applications, we study in detail the dependence of the quality of a model application on the data quality. The quality is manipulated in a controlled way by adding noise to the data. A real-life situation may be video signals with corrupted sensors, e.g. by sand or water on the optics. Using the described scenario, we can determine scaling behavior over three decades of noise.}

%Articles submitted to Chaos should additionally contain a lead paragraph. This paragraph, which will be highlighted in the journal in boldface type, essentially advertises the main points of the article. It must describe in terms accessible to the nonspecialist reader the context and significance of the research and the importance of the results. The editors pay special attention to the clarity of this lead paragraph and may suggest edits.

\section{Introduction}

\label{sec:introduction}

Real data contain corruptions from multiple sources, ranging from inaccuracies in sensors, environmental impacts during data acquisition to noise in the data generating process itself. Such imperfections influence the quality of technology that is based on data, and physical progress. In physics, the empirical validation of models has a long-standing and well understood foundation \cite{popper}.
We focus on Artificial Neural Networks (ANNs) which have empiricism built into the development of corresponding models by their train-test-validate pipeline to develop models to infer properties of data \cite{Goodfellow-et-al-2016}. After establishing the models based on data, they are used, which is referred to as \textit{inference}.
In real-world applications, the data quality may change during data acquisition for said inference. Therefore, it is of utmost importance to understand how much corruptions a given ML model can sustain to achieve a certain purpose.

Consequently, we study the following  basic question: how does the model quality, \textit{accuracy}, or \textit{score}, respectively,  change with noise on data? I.e. we consider the transition from cognition to ignorance, i.e. from perfect prediction to no ability to predict. Asymptotically, a very large model system, and infinitely many, perfect, and noiseless data the model will show perfect prediction accuracy. However, real-world systems are finite and we have to consider two scenarios: i) a reduced, or finite model size, and ii) noisy data. We do not discuss the third important parameter, namely the limited number of data, here.

There are practical implications: the choice of the model size, or dimensions, is crucial for any application in terms of resources, i.e. cost. For a required accuracy, smaller than optimal sizes may be acceptable and save costs. This is particularly important for applications on hardware with limitations, e.g. on edge devices with small storage capacity, bandwidth, and relatively slow compute units.

Typical limitations of sensors are: noise from environment, noise from the sensor itself, and noise in Bits and Bytes on digital devices. For spatial measurements, e.g. images, radar or lidar signals, the resolution is another limitation. The same holds, of course, for temporal signals, where the sampling rate is given by the sensor device. In fact, time and space resolution is often coupled, e.g. for the above mentioned radar signals. Here, we do not go into detail of a particular measurement process, rather want to study the general mechanism how the prediction of a model changes with measurement noise. Consequently, we control noise by a controlled, artificial modification of signals which are of good original quality. To not deviate from the fundamental aspect, we limit the investigation to Gaussian noise, as used for stable diffusion \cite{rombach2022high,SohlDickstein2015}

There are numerous applications where the above scenario plays a role:
autonomously driving cars may need to "understand" when they are no longer able to drive if a sensor is broken or the environment is hostile (rain, snow, ...), similar to humans who stop in critical situations; a LLM may need to detect distorted spoken input, e.g. if a microphone is not working, and respond accordingly; unsupervised learning models may have to have mechanisms to respond to corrupted data to avoid the "garbage in - garbage out" scenario.

Given the described transition, we think it is fair to sketch an analogy to known and well understood physics of large systems, namely statistical mechanics. We borrow corresponding concepts \cite{huang2009introduction} to answer the mentioned questions, our subject of study is a well understood scenario, namely image classification. As a first step, we clarify how the cognition transition from high accuracy to inaccurate prediction takes place. For a large system, we expect a sharp transition with increasing noise, where the steepness of the transition increases with increasing model size \cite{icaart24}, and is washed out with decreasing system size.
I.e. we study noise, but as well finite-size efects \cite{Binder}.
Statistical mechanics with respect to machine learning has been investigated in the last decades \citep{Bahri2020,Carleo2019,Mezard2009} with strong reference to early work on non-equilibrium thermodynamics of learning \citep{Gardner1988,Sompolinsky1990,gyorgyi_first-order_1990,Seung1992,Watkin1993}, and more recent advances by \citet{Biehl2007c}, and \citet{advani_statistical_2013}.

We observe that the transition happens at a critical noise which can be identified for the models investigated very accurately. The second step concerns a quantitative study using a scaling approach. Our original motivation was the search for universality in deep learning models, where one obtains the same scaling behavior for models which are similarly built, or in the same universality class. Unfortunately, we could not drive our study to a large enough variety of models due to computational limits.  We find, however, indications for an understanding of asymptotic behavior of models when close to the critical point. The emphasis here is not on replicating real-world noise but on exploiting the statistical properties of artificially introduced, controlled noise. De-noising strategies and model architectures are reviewed by \citet{TIAN}, with notable applications found in \citet{Smilkov2017,Koziarski2017}.

Apart from the fundamental value in gaining improved understanding, there is a very important practical aspect: the dimensioning of models while considering data corruption scenarios. The introduction of  university classes of models would provide a strong tool to optimize similar models, e.g. by selecting the range for hyperparameter studies.

The paper is structured as follows: After this section, we provide a more detailed account of the background in Sec.\ref{sec:background}. We then present the method and data we use. We then present the results of our study. Ultimately, we discuss the possible implications of our findings and give further research perspectives.

\section{Background}

\label{sec:background}
There are two sides of the described transition, on one hand the practical aspects with respect to AI, on the other hand the more fundamental considerations in relation with statistical mechanics. In the following, we first relate our work to Deep Learning application and then draw the analogy to phase transitions and statistical mechanics.

Generally, information destruction in data  reduces deep learning model performance, as reviewed by \citet{drenkow2021systematic}.  However, it is noteworthy that, for successful learning, overall information may be reduced, as long as the necessary information is kept, e.g. when reducing image size for classification \cite{rombach2022high}. Here, we chose image recognition applications, which are very important in applications, have well studied models and curated data sets \cite{DL4visionsystems}. In
image recognition applications, noise on the images may be due to distortion of the sensors, or due to technical limitations, like storage or transmission errors. In many cases, it can be approximated as additive Gaussian noise \citep{BONCELET2009143}, where an artificially generated noise term is added to images. This additive procedure has recently gained attention in the field of generative, diffusion-based models \citep{SohlDickstein2015,Ho2020}.
We will use this approach later to add noise to images in a controlled way.

ML model performance is quantitatively assessed using metrics that statistically summarize the inference on all  data points in train, test, and validation datasets. The split into the said three data sets follows the empirical approach using statistical arguments for good sampling \cite{statisticalLearning}: by training models are iterated until they approximate well the given data, by testing the parametrization is applied to an independent test data set, to avoid overfitting and achieve good performance. This train-test cycle is iterated until convergence is achieved; statistical validity can be improved, e.g. by crossvalidation \cite{statisticalLearning}. Eventually, the performance is validated by the until then unused validation data set.

The performance, or score, is a macroscopic quantity which summarizes measures the response of the neural network to an external input, the data.
In statistical mechanics, the analogy is the macroscopic order parameter which measures response of a material. A classical example are spins in a magnetic material as analogous to the neural network, and magnetization as order parameter. The data, in this analogy correspond to an external field that acts on the magnetic material, or the spins, respectively.

In classification tasks the analogy to spin systems works as follows: the accuracy of a classification model describes the average over a binary state variable, that is typically $1$ for correct classification on a data point and either $-1$ or $0$ for an incorrect classification. In analogy to statistical mechanics data are an external force that let the system, or model, change until the order parameters, or internal configuration of the system are optimized accordingly.

We want to study noise - in physical systems, noise is typically of thermal origin. Temperature drives phase transitions in order parameters of ensembles of microscopic constituents in dependence of noise as a collective nonlinear response which optimizes the material's free energy, for an introduction cf. \cite{huang2009introduction}. The transition happens from a nonzero value of the quantity under consideration, e.g. magnetization or performance at a critical value of the varied parameter, e.g. temperature or noise. Such a phase transition is characterized by a change in one or more order parameters or their derivatives. Such transitions may be driven by  external fields, temperature, or internal parameter changes.
They occur across basically all natural sciences and beyond; general aspects on phase transitions in non-physical systems are mentioned for biology \citep{Heffern2021}, computer science \citep{MARTIN2001}, social science \citep{Perc2017}, and especially in the dynamics of learning \citep{Seung1992,Watkin1993,Carleo2019,Bahri2020}, to name but a few examples. In turn, ML models are used to identify phase transitions in several physical systems, e.g. in quantum mechanics \citep{Rem2019}, complex networks \citep{Ni2019}, and condensed matter \citep{Carrasquilla2017}. More general approaches to detecting phase transitions using Machine Learning are found in \citet{Canabarro2019,GIANNETTI2019114639,Suchsland2018}. This is, however, not the topic of this study.

Continuous phase transitions are characterized quantitatively by critical exponents which describe a power law behavior of a physical quantity, like the mentioned magnetization, when approaching the critical temperature. When the critical exponents for two processes considered are equal, the transition is said to belong to the same universality class. This is a strong statement and allows to conclude properties of material without detailed studies. That motivates us to study scaling and exponents and criticality in neural networks.

Scaling Laws describe how a characteristic or quantity changes  by approaching  a critical point $x_c$. This "approaching", can be formally described as a limit process  $\lim_{\epsilon \to 0} \|Q(x_c+\epsilon) - Q(x_c)\|$, with $Q$ the quantity under consideration. The limit process extends over several orders of magnitude of the independent variable, i.e. several scales, hence the name. That can often be best visualized and understood using logarithmic scales to reveal power laws.
Real-world systems display finite-size effects, as they do not reach the thermodynamic limit in size, or have infinite time to converge. A sharp, or well-defined phase transition is then smoothed out about the critical point, with the concrete properties depending on the system. However, general behavior when approaching the transition point is again governed by universal scaling laws, cf. \cite{Binder}.

In deep learning, scaling laws often refer to how model performance scales with data size, model parameters, or computational resources, said scaling is analogous to the scaling laws related to finite-size effects.
Such finite-size effects occur, of course in deep learning models: the performance of machine learning (ML) models is intrinsically linked to the quality and quantity of data used in training and inference \citep{cortes1994limits} and the number of parameters in a model, or its size, respectively. The efficiency in terms of computational cost and speed is related mainly to, again,  dataset and model size \citep{ALJARRAH201587}. With the advent of foundational models, neural scaling laws have gained significance, i.e., the scaling of model performance with the amount of training data and model size  \citep{Kaplan2020}. For practitioners it is a crucial interest to gauge the expected improvement in model performance depending on improved data quality and amount of data. In other word, how much performance has a practitioner to trade for a significant cost reduction and sustainability improvement.

In any application, included ML, one is faced with finite system sizes. This is not ideal for the so-called thermodynamic limit, where one considers infinite system size. This limit is particularly interesting, because interacting systems with repeating or similar units exhibit generic asymptotic behavior, described by statistical mechanics.

Any ML system has a finite size, and consequently we focus on finite-size scaling that describes the corrections to this idealized system behavior. With  increasing system size the following holds: a property $F(L)$, with  system size $L$ is defined as $F(L) \propto L^{-\lambda}f((T-T_C)L^{1/\nu})$ where $\lambda$, $\nu$ are critical exponents and f is a scaling function.
The definition of $L$ may be very coarse, e.g. the number of cells in a network, no matter what type and function.

Universality in deep learning refers to the idea that systems with different microscopic details  exhibit the same macroscopic behavior near critical points. Universality implies that neural networks can behave similarly across different architectures and tasks when they are in the same universality class. This contains enormous practical potential, since many expensive data and model training runs could be avoided when the behavior of training, testing and validation is known through the universality class.

\section{Methods and Data}

\label{sec:method}

As mentioned above, we investigate finite-size scaling in AI by image recognition. In particular, we focus on the degradation of model accuracy with increasing added noise to data. There are two ways to study this dependence: train, test, and validate the accuracy on corresponding data sets, or train and test a model with a good data set to obtain the parameters of a neural network; the resulting  model is called pretrained. The subsequent application to noisy data corresponds to a more realistic situation. Henceforth we can utilize pretrained models and  we vary the model size in the presence of noise in the data, such that we cover the above described regimes of high and low signal/noise ratio, or unimportant/dominant noise, respectively.
Some extensive studies regarding the scaling of accuracy with respect to model size have already been presented by \cite{Kaplan2020}, here we combine it with the said noise study.

In this section, we firstly and briefly describe the data used and how we noisify them; second, we describe the AI experiments we run for our study.

\subsection{Data and Noise Application}
\label{sec:noise}

For our study, we required a well-studied and understood data set to avoid pitfalls with unexpected data peculiarities. For image recognition, the probably best understood data set is Imagenet\cite{imagenet_cvpr09} with its descendants. Consequently, we used models have been pre-trained on the ImageNet21k dataset and finetuned on ImageNet2012, but also trained smaller models from scratch on CIFAR100 \cite{krizhevsky2009learning}.
In particular, we make use of the EfficientNet architecture and the scaling scheme as introduced by \citet{tan-le-2019}.

The modeling of realistic noise depends on the concrete situation, e.g. if sensor data are concerned the sensor characteristics is relevant. This holds for photographs, videos, but as well for other sources, like sound or radar waves. Here, we focus on images. Noise can be additive or multiplicative, and be subject to very complex generation mechanisms \citep{Kampen1992}.

We introduce corruptions to the data with a peculiar scheme. In the regime of small noise (signal dominates noise intensity) one usually simply adds Gaussian noise with 0 mean and small variance. However, in the regime of large noise, this scheme introduces unwanted effects as the domain of greyscale values is bounded and values are typically discrete. To mitigate these limitations, we use the noise adding scheme as developed in the context of Denoising Diffusion Probabilistic Models \cite{SohlDickstein2015}, \cite{Ho2020}
in which the original (unknown) data distribution is transformed into a well-known target distribution, in our case again a Gaussian distribution.
The transformation from a distribution of image data to well-known noise-distributions has been presented in Denoising Diffusion Probabilistic Models (DDPM) \cite{SohlDickstein2015}. In training these generative models, the forward process is that of corrupting images with noise, while the reverse process consists of denoising these images again. The reverse process is the objective of training the generative models. In our experiments, however, we only use the forward process and refer to it as perturbation process. In this section we give an outline of the image corrupting forward process as presented in \cite{SohlDickstein2015}, \cite{Ho2020} which we use to perturb our data with additive noise in all experiments.

An initial data point (e.g., an image) is gradually corrupted by (multivariate) Gaussian noise over a series of time-like steps, which we use as an indicator of noise intensity in our data.

\subsubsection{Perturbation by Noise}
The perturbation process involves a sequence of latent variables $\{x_0,x_1,...,x_T\} \in \mathbb{R}^{m\times n}$ where $x_0$ is the original data (image with $m\times n$ pixels\footnote{For simplicity we write $\mathbb{R}^{mn}$ since $\mathbb{R}^{mn}$ is isomorphic to $\mathbb{R}^{m \times n}$.}) and $x_T$ is the final noise-corrupted image. The process is defined as a Markov chain \cite{MeynTweedie} where each $x_t$ is generated from $x_{t-1}$ by adding Gaussian noise. We give the evolution equation for the image distribution $q(x_0)$ and conditional distributions $q(x_t|x_{t-1})$ at a given step $t \in \{0,...,T\}$. This is done iteratively according to the equation:

\begin{equation}
    q(x_t|x_{t-1})=\mathcal{N}(x_t;1-\beta_t x_{t-1},\beta_t\mathbb{1})
\end{equation}

Here:
\begin{itemize}
    \item $q(x_t|x_{t-1})$ is the conditional probability distribution for $x_t$ given $x_{t-1}$.
    \item $\beta_t$ is a variance or noise schedule, which is a small positive scalar controlling the amount of noise added at each step.
    \item $\mathcal{N}(\cdot;\mu,\sigma^2\mathbb{1})$ denotes a multivariate Gaussian distribution with mean $\mu \in \mathbb{R}^{mn}$ and covariance $\sigma^2\mathbb{1} \in \mathbb{R}^{mn\times mn}$.
    \item $\mathbb{1}$ is the unit matrix.
\end{itemize}

\subsubsection{Noise Schedule}

The noise schedule $\{\beta_t\}_{t=0,...,T}$ is often designed to increase monotonically over time, starting from a small value (minimal noise added in the first steps) to a larger value (significant noise in the final steps). We use a linear noise schedule given by

\begin{equation}
    \beta_t = \beta(t) = 1 \cdot 10^{-4} + 4\cdot 10^{-6}\cdot t
\end{equation}

and $T=5000$ noise steps. Note that despite the small size of the variance $\beta_t$ at each step, the repeated addition of noise still yields a noise dominated image for $t\to T$.

\subsubsection{Marginal Distribution of $x_t$}

The perturbation process allows us to directly sample $x_t$ from $x_0$ (the original image) without sequentially sampling every intermediate step, thanks to the properties of Gaussian distributions. The marginal distribution of $x_t$ can be derived as:

\begin{equation}
    q(x_t|x_0)=\mathcal{N}(x_t;\bar{\alpha}_tx_0,(1-\bar{\alpha}_t)I)
\end{equation}

where $\bar{\alpha}$ is defined as:

\begin{equation}
\bar{\alpha}_t=\prod_{s=0}^t(1-\beta_s)
\end{equation}

This shows that $x_t$ is a noisy version of the original image $x_0$ with noise level controlled by $\beta_t$ through $\bar{\alpha}_t$.

The choice of variance schedule $\{\beta_t\}_{t=0,...,T}$ needs to ensure that $\bar{\alpha}_t \rightarrow 0$ for $t \to T$ so that we approximate a Gaussian distribution of zero mean and unit variance.

Note that $\sigma := \sqrt{(1-\bar{\alpha}_t)}$ defines the standard deviation in the sampling scheme, while $\mu := \sqrt{\bar{\alpha}_t}$ scales the mean value of the distribution.

\subsubsection{Important Properties}

\begin{itemize}
    \item Independence: The forward process defines a Markov chain, meaning the state $x_t$ only depends on the immediate previous state $x_{t-1}$.
    \item Gaussianity: At each step, the noise added is Gaussian, and due to the linearity of the Gaussian distribution, the resulting distribution remains Gaussian.
    \item Analytic Marginals: The marginal distribution $q(x_t|x_0)$ can be computed analytically, which is useful for efficient training and sampling.
\end{itemize}

%\subsubsection{Intermediate Sampling}

For a specific time step $t$, given the original image $x_0$, you can directly sample an intermediate noisy version using:

\begin{equation}
\label{eq:noiseSchedule}
x_t=\bar{\alpha}_tx_0+1-\bar{\alpha}_t\epsilon
\end{equation}

where $\epsilon \sim \mathcal{N}(0,I)$ is multivariate Gaussian noise.

\subsection{Transition Study}
We want to investigate the robustness of artificial neural networks (ANNs) by varying the number of hidden units, and noise intensities in the data. The goal is to understand the transition from optimal model accuracy to baseline accuracy, i.e. randomly guessing the label for each image.

As for our choice of the data set we choose equally well understood model, namely   EfficientNet \cite{tan-le-2019} , both in the sizes from the original publication and customly trained smaller models. From universality theory we expect some dependence of the scaling on the underlying symmetries in data and model. The number of trainable parameters for each model are given in table \ref{tab:model_size}.

We use pretrained weights on Imagenet21k that are finetuned on ImageNet2012 \cite{imagenet_cvpr09} for EffiecientNet $B0,...,B7$. Weights were taken from keras applications \cite{chollet2015keras}. For details on the model training refer to \cite{tan-le-2019}. These models were evaluated on the corrupted imagenette validation dataset \cite{Howard_Imagenette_2019}, a subset of ImageNet2012. Since the scaling is not very expressed for these large models, we additionally train smaller EfficientNet models from scratch on CIFAR100 \cite{krizhevsky2009learning} and evaluate on corrupted CIFAR100 data.
%TODO Extra table with our model sizes
The data corruption follows \cite{SohlDickstein2015} but was implemented independently.

\subsubsection{Evaluation}

Our main experiment consists in adding noise to the test data and recording the accuracy as given by the formula

\begin{equation}
	acc(y, \hat{y}) = \frac{1}{n_{samples}}\sum \mathbf{1}(\hat{y}_i=y_i)
\end{equation}

where $y$ is the predicted label, $\hat{y}$ the true label, and $\mathbf{1}(a,b)=1 \ if \ a=b, \ else \ 0$ is the indicator function.

Other metrics, namely F1 score, precision, and recall on have been recorded as well, but did not yield additional insights. To specifically measure robustness, we measured the performance of each model as a function of noise intensity.

We use the full validation set of $50000$ images from ImageNet2012 in a $384 \times 384$ resolution for the larger models. Noise is added batch-wise in batches of $512$ images so as to marginalize the inherent stochasticity in noise generation.
The smaller models are evaluated on CIFAR100 validation set in varying batch sizes.
%TODO Adjust for custom models and CIFAR100

%EfficientNet models are imported from keras applications \cite{chollet2015keras} and provided weights were used for the models $B0,...,B7$.

%For dataset loading we use tensorflow datasets \cite{TFDS} with a batch size of 512 on the validation split. %TODO This has been mentioned before

\begin{table}
\centering
\begin{tabular}{| c | c | c |}
\hline
Model & \#Parameters &FLOPS \\
\hline
EfficientNetB0 & 5.3 & 0.39 \\
\hline
EfficientNetB1 & 7.8 & 0.70 \\
\hline
EfficientNetB2 & 9.2 & 1.0 \\
\hline
EfficientNetB3 & 12 & 1.8 \\
\hline
EfficientNetB4 & 19 & 4.2 \\
\hline
EfficientNetB5 & 30 & 9.9 \\
\hline
EfficientNetB6 & 43 & 19 \\
\hline
EfficientNetB7 & 66 & 37 \\
\hline
\end{tabular}
\caption{Table that shows the number of trainable parameters and FLOPS for each of the used models, see \cite{tan-le-2019}.}
\label{tab:model_size}
\end{table}

\subsection{Determining scaling coefficients}

As described above, in the context of phase transitions, the idea of scaling refers to the approach of a critical point by smaller and smaller "scales", in a logarithmic sense. E.g. by considering the distance in steps $-1,-2,-3,...$ of the exponent of 10: $10^{-1}$, $10^{-2}$, $10^{-3}$, and so on.
Typically, one observes power laws when displaying an observed quantitiy, best recognized in a double logarithmic graph. One speaks of power law scaling \cite{STANLEY200060,Lovejoy2023}.
Here, we apply the ideas to observe the behaviour of model accuracy when approaching the critical point of cognition transition from a noiseless situation to the scenario with additional noise.

\begin{figure}
    \centering
    \includegraphics[width=\linewidth]{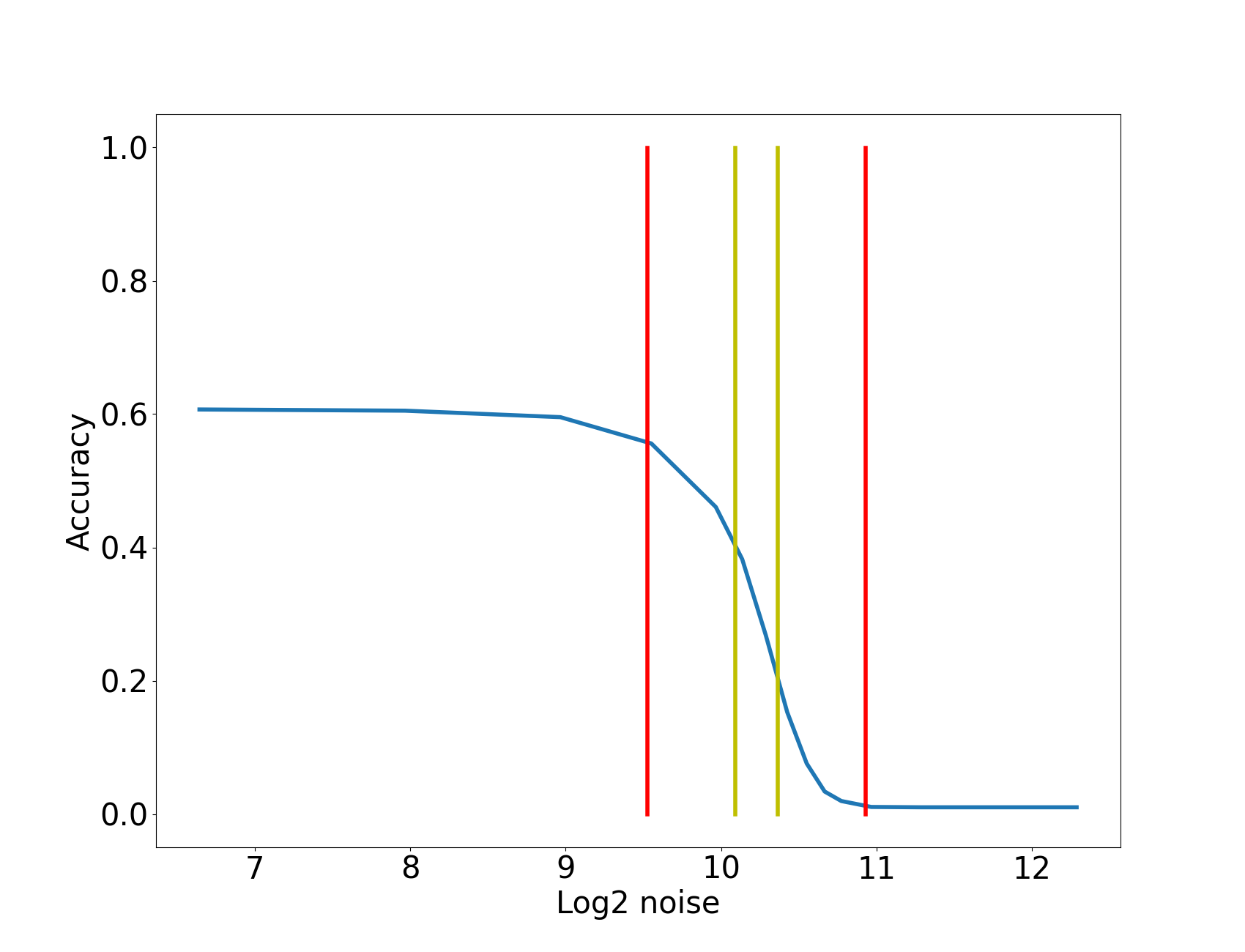}
    \caption{Visualization of the level set approach to produce a series of nested intervals on a logarithmic scale for levels $2^{-1}$ \textit{(red)} and $2^{-5}$ \textit{(yellow)}. }
    \label{fig:levelset}
\end{figure}

The focus of our study lies on finite size effects - they do not only induce a scaling of model accuracy, but also a scaling of the transition's width. This width narrows with more and more network units, or nodes, and the question is if it follows the expected asymptotic power-law behaviour. In the  (thermodynamic) limit of infinite model size the width , the finite size of the models induces a continuous transition, the scaling coefficient of the width of which can be calculated using the level set method as introduced by \citep{Evans1991}, \cite{Evans1992}. That is, we define levels of the accuracy  and determine the according noise value. To achieve the above described scaling, we enclose the critical point and produce nested intervals. These nested intervals can be chosen with logarithmic levels of accuracy. Eventually, we display these levels vs the model size to understand finite size scaling.
The procedure is visualized in Fig.~\ref{fig:levelset}.

\section{Results}
\label{sec:results}

As described above, we study the dependence of the width of the cognition transition with increasing system size. To recall - the idea behind is to obtain a guideline on ANN dimension for a given data characteristics. Thus, our study aims at determining scaling laws of model performance with respect to size under the assumption that the cognition transition with respect to noise in the data exists and follows statistical laws. In statistical mechanics, the asymptotic limit $N_{net}\mapsto \infty$ is approached, such that a discontinuous transition is observed. Here, we have a closer look at finite-size effects.

First, we inspect the existence of the promised cognition transition for one model, namely the EfficientNet model class. The details of the models is found in \cite{tan-le-2019}, the essential parameter for us is the total number of parameters is given in Table~\ref{tab:model_size}.
Let us inspect that figure: we clearly identify a sharp bend towards near zero accuracy with increasing noise in the range 1000 - 2000 (remember that we plot against the noise step as used in stable diffusion models, and not against the noise level or the signal-to-noise ratio).
Here, the signal-to-noise ratio is so small that there is no more information left and classification fails. What is interesting, though, is that there is a universal way this change of cognition capability changes with noise scale.

\begin{figure}[!h]
	\centering
	\includegraphics[width = \linewidth]{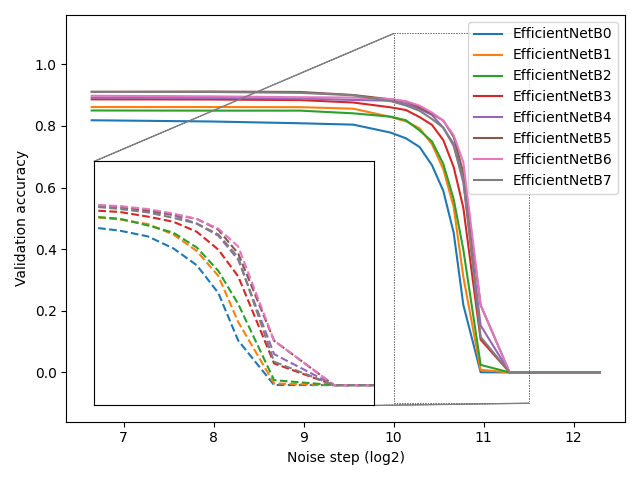}
	\caption{EfficientNetB{0...7} accuracy vs.  noise intensity. Noise is measured in step number of the diffusion increase, cf. Eq.~\ref{eq:noiseSchedule}. We observe a sharp transition for any network size. The inset shows a magnification of the transition region $log(step) \in (10,12)$, one has a unique critical value of $t_c = 1845.934$ for all models. The model size varies over approximately one decade. For a scaling study, this is not adequate, rather one should study 3-4 decades to obtain a meaningful power law. }
	\label{fig:EffNetB0-7}
\end{figure}

Our main goal is to understand the cognition transition in accuracy with respect to noise  is shown in Fig \ref{fig:EffNetB0-7} for EfficientNets B0-B7.
Generally, larger models perform better and show a sharper transition. Both effects are expected from finite size scaling theory and known scaling laws for EfficientNet performance.
Notably the critical noise is the same: the unique critical noise value for all EfficientNet models is $t \approx 2000$, corresponding to $log(\sigma_c) \approx 11$.

The next step is to apply the level set method to the transition curves. First, however, we extend the range for our model sizes studied. We use EfficientNet-compatible sizes and repeat the study presented in Fig.~\ref{fig:EffNetB0-7}. Then we choose levels $(a_{low,i},a_{up,i})$ equally distant from 0.5, e.g. $(0.1,0.9)$, $(0.2,0.8)$ and construct a typical width of the transition $\Delta = |\beta(a_{up}) - \beta(a_{low})|$ for each iteration step $i$. Why use different level sizes? It is a means to ensure that we have a converged value for Delta when approaching $|a_{up} - a_{low}|\to 0$, the thermodynamic limit one would achieve for infinitely large systems and an infinite number of data.

\begin{figure}[!h]
	\centering
	\includegraphics[width = \linewidth]{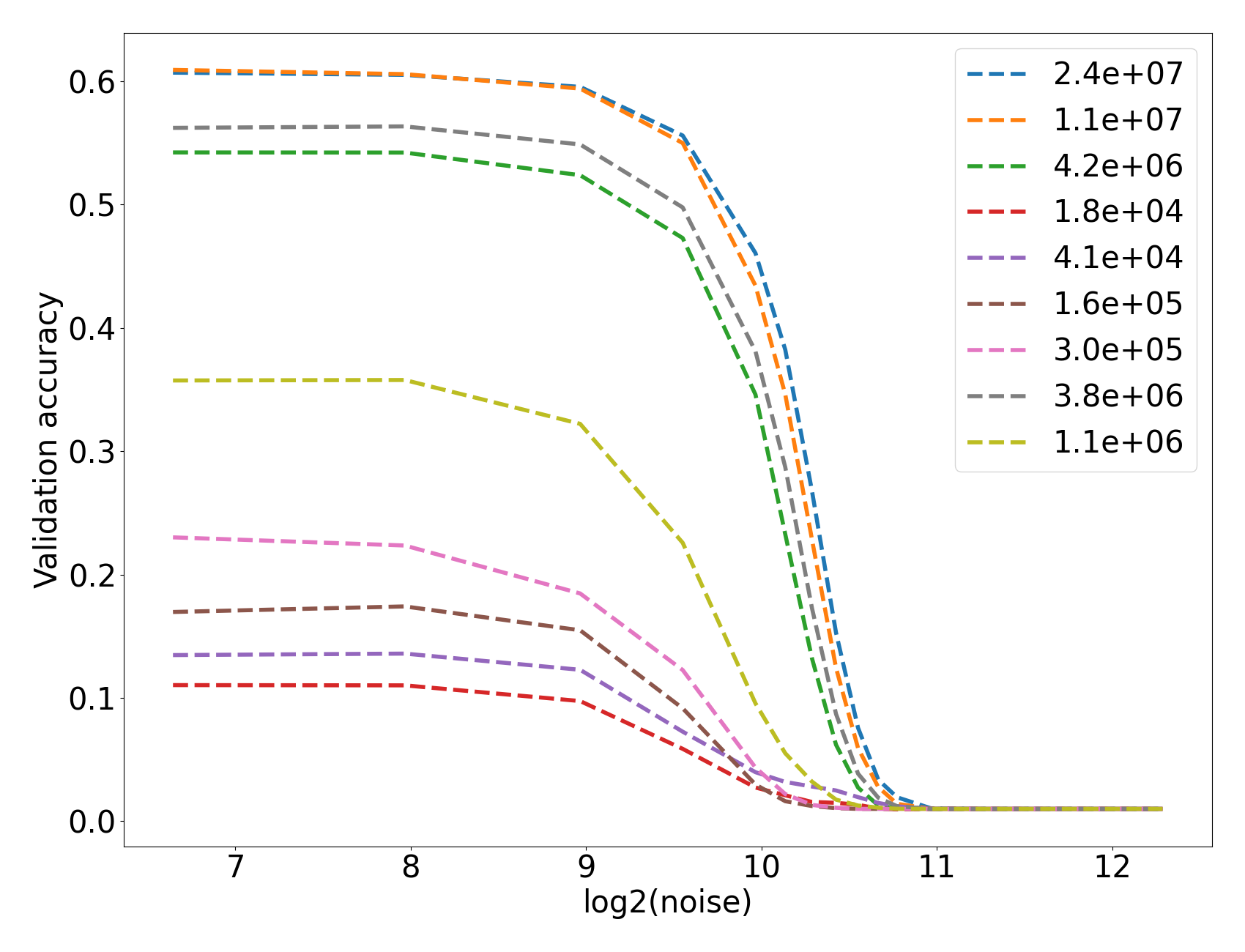}
	\caption{Scaling of validation accuracy of EfficientNet type models trained on CIFAR100 and evaluated on data with added noise.}
	\label{fig:scaling_EffNet}
\end{figure}

The transitions for a range of 3 decades are displayed in Fig.~\ref{fig:scaling_EffNet}. Since the models are very small in part, starting with $10000$, the accuracy is very bad, accordingly. However, for our study, this does not play a major role, in fact it may reveal the strength of our approach: we can obtain a typical behaviour, even for models that are of no practical use. With their scaling, one can subsequently conclude a useful model size for the problem at hand.
Let us inspect of this hypothesis is confirmed.

\begin{figure}[!h]
	\centering
	\includegraphics[width = \linewidth]{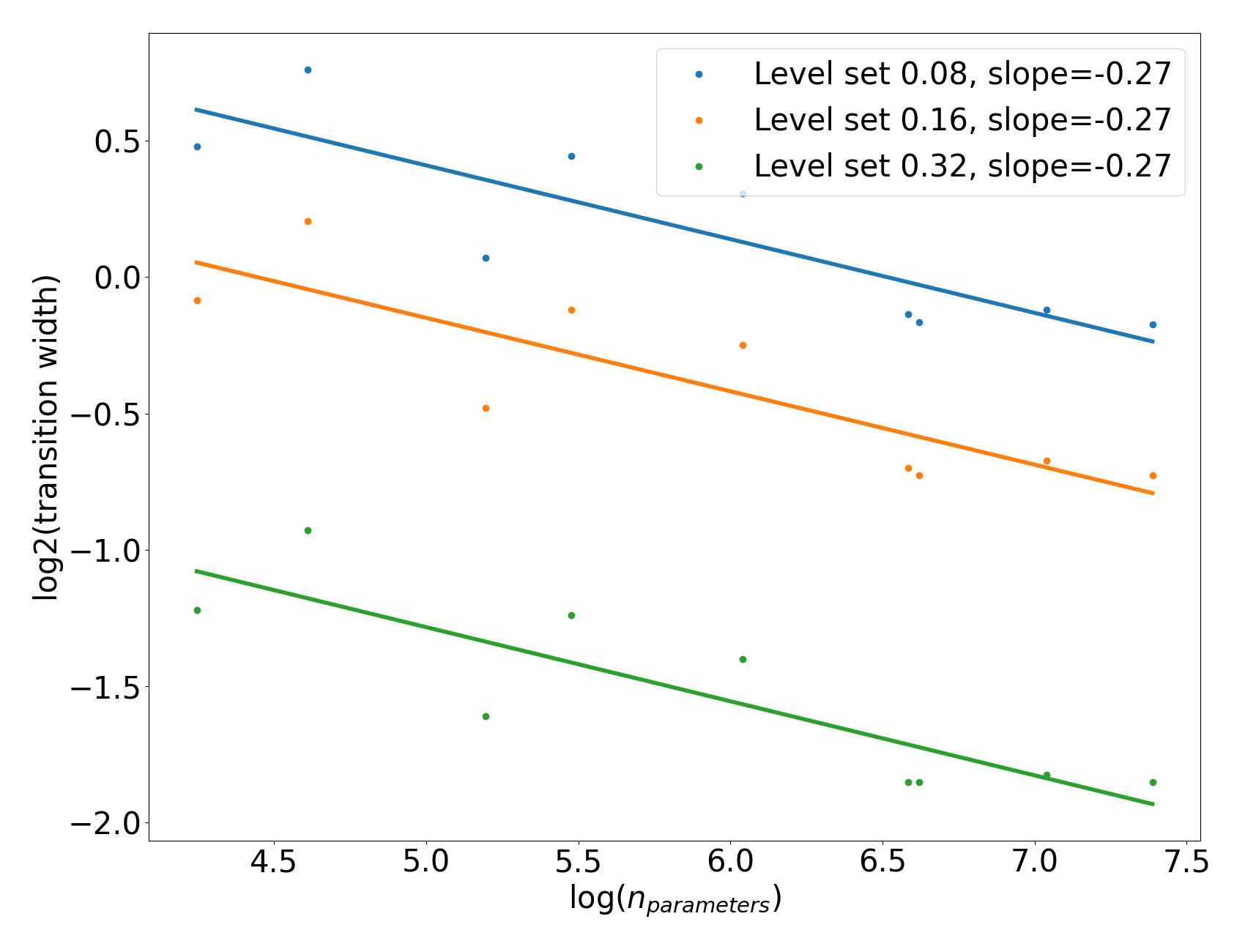}%fig/TransitionWidths_ModelSize_effnet_cifar100.png}
	\caption{Width of the cognition transition for EfficientNet type models trained on CIFAR100 and evaluated on data with added noise for different level sets.}
	\label{fig:scaling_final}
\end{figure}

Scaling and scaling exponents, if existent are obtained in the way described above, consequently we display the logarithm of the transition width $\Delta_{acc} = |a_{up} - a_{low}|$ vs the logarithm of the system size $N = \#\mathrm{params}$ in Fig.~\ref{fig:scaling_final}. We do so for different level sets $i$ to eliminate aleatoric effects of the level set. It turns out that we obtain a power law with slope $0.27\simeq 0.3$. This holds across three decades.

The different level sets chosen behave in the same way, even points above and below the power-law fit show systematically the same behaviour. Let us comment on these deviations: it is typical for scaling that, with decreasing resolution, the deviations from the power law become smaller and smaller, since the asymptotic behaviour is approximated better and better. I.e. for small size deviations are large, for large sizes, they are small. This tendency can be clearly seen in Fig.\ref{fig:scaling_final}. In fact, it is surprising that the variation is not larger, given the very small model size, in comparison of the many data to be trained and classified!

\section{Discussion and Conclusion}

\label{sec:conclusion}

In this study, we demonstrate that principles of statistical mechanics hold for AI. We have focused on the cognition transition as a transition between two phases - cognitive or not - characterized by the ability of the models to accurately recognize images. Based on our studies, we expect that such transitions are a general feature of deep learning, where the basic condition is that models  consist of many similar units such that it makes sense to apply statistical considerations and consider finite size effects when the system size approaches the thermodynamic limit.

The first issue we tackled was the clarification if a cognition transition can be found for data subject to Gaussian noise. Then, we focused on finite-size effects, and in particular the asymptotic behaviour of the model's accuracy when approaching the critical noise level.
The mathematical tools used by us are applied to numerical studies, i.e. we run analyses with models of different size, and study if scaling can be observed in the AI models. As a result, we can verify scaling behavior for a family of models: EfficientNet. That means, we have a power law in the decrease of accuracy when approaching the critical point from the cognition side.

Our hypothesis is that this implies universal behavior of all models in the same class. Whereas that sounds great, we cannot yet pin down which other models that may be in general. From theory, one expects models with similar structure and constituents to behave similarly.
To verify that in more detail, we need to run a large study of possibly very diverse models for image classification.

The practical use lies in the estimation of an optimal  model size, used for a certain task. In our scenario, this concerns data with a certain noise, or in other words signal to noise ratio. Alternatively, one may discuss the transition in terms of information contained in the images. So, once determined the scaling for small systems, an extrapolation allows the determination of how well a model may asymptotically perform with decreasing or increasing noise, respectively.

Noise on data is an important practical phenomenon. Our approach may be used for other cognitive transitions, e.g. for studies if the transition point may shift for various models. In addition, other quantities may be studied which are subject to fluctuations or to finite-size effects.

Concluding, we may say that these first systematic results look very promising and have on one hand practical applications, on the other hand may help to understand better the mechanics of large models. Given nowadays interest in ever larger models, we are tempted to say that this is a very important use, where know-how transfer from physics and mathematics can be of great help in the field of AI.

\section*{Acknowledgement}
\label{sec:acknowledgement}

We acknowledge fruitful discussions with M. Schultz, K. Wiesner, and T. Ersoy on the transition mechanism. MA was supported by Bundesministerium für Bildung und Forschung (BMBF), project CeCas, Grant No. 16ME0810, TS was supported by the Bundesministerium für Klimaschutz, Umwelt, Energie, Mobilität, Innovation und Technologie (BMK), within the project KI:STE, Grant No. 67KI2043.


%\nocite{*}
%\bibliography{literature}% Produces the bibliography via BibTeX.

%\bibliography{literature}

%\section*{Acknowledgement}
%We acknowledge fruitful discussions with F. Emmerich. This work is supported by the KI:STE project, grant no. 67KI2043 by the German Ministry for Ecology. MA acknowledges support by the national German Federal Ministry of Education and Research under the grant ME160810 and 16ME4053.
\end{document}